\documentclass[12pt]{iopart}

\usepackage{amsfonts}
\usepackage{amssymb}
\usepackage{booktabs}
\usepackage[shortlabels]{enumitem}
\usepackage[T1]{fontenc}
\usepackage{graphicx}
\usepackage[colorlinks=true, linkcolor=blue, urlcolor=blue, citecolor=blue]{hyperref}
\usepackage{cleveref} 
\usepackage{multirow}
\usepackage{siunitx}
\usepackage{xcolor}
\usepackage{xspace}
\usepackage{orcidlink}

\usepackage[
    backend=biber,
    style=numeric-comp,
    sorting=none,
    mincitenames=1,
    maxcitenames=2,
]{biblatex} 
\newcommand{\A}{\mathcal{A}}

\newcommand{\avg}[1]{\langle #1 \rangle}
\newcommand{\chtime}[1]{\tau_{\textrm{#1}}}
\newcommand{\dc}{\textrm{dc}}

\newcommand{\eah}{\texttt{GCT}}
\newcommand{\F}{\mathcal{F}}

\newcommand{\Sn}{S_{\textrm{n}}}
\newcommand{\snrSqAv}{\avg{\rho^2}}
\newcommand{\snrSqAvASky}{\snrSqAv_{\vec{n}, \cos\iota, \psi}}
\newcommand{\T}[1]{T_{\textrm{#1}}}

\newcommand{\weave}{\texttt{Weave}}
\newcommand{\team}[1]{\texttt{#1}}

\newcommand{\pfp}{p_{\mathrm{FP}}}

\newcommand{\pfd}{p_{\mathrm{FD}}}

\newcommand{\C}{\mathcal{C}}

\DeclareSIUnit{\day}{\textrm{d}}
\DeclareSIUnit{\month}{\textrm{month}}
\DeclareSIUnit{\persqrthz}{\hertz\tothe{-1/2}} 
\DeclareSIUnit{\rad}{rad}

\begin{document}

\title{Learning to detect continuous gravitational waves: an open data-analysis competition}

\author{
    R Tenorio$^{1,2,3}$\,\orcidlink{0000-0002-3582-2587},
    M J Williams$^{4,5}$\,\orcidlink{0000-0003-2198-2974},
    J Bayley$^{5}$\,\orcidlink{0000-0003-2306-4106},
    C Messenger$^{5}$\,\orcidlink{0000-0001-7488-5022},
    M Demkin$^{6}$,
    W Reade$^{6}$,
    J Koda$^{7}$,
    Y Yamakawa$^{7}$,
    T Yamaguchi$^{7}$,
    K Abe$^{7}$,
    C Achard$^{7}$,
    H S T Bukhari$^{8}$,   
    M V Shugaev$^{9}$,
    G Sokolov$^{7}$,
    H Yoshihara$^{10, 11}$,
    V Debout$^{12}$,
    S Goulet$^{12}$,
    J-L Tastet$^{13, 14}$,
    I Timiryasov$^{15}$,
    O Ruchayskiy$^{15}$, 
    D Kanonik$^{7}$\,\orcidlink{0000-0000-0000-0000},
    S Seferbekov$^{7}$,
    S Saito$^{7}$,
    R Sato$^{7}$,
    S Segawa$^{7}$,
    A Zhyvalkouski$^{7}$,
    Y Uchida$^{7}$\,\orcidlink{0000-0002-6932-1465},
    S Yokoi$^{7}$,
    A Sayed$^{7}$,
    R-Q Xing$^{7}$,
    I Yamashita$^{7}$,
    Z Wang$^{7}$
}
\address{$^1$ Dipartimento di Fisica ``G. Occhialini'', Universit\`a degli Studi di Milano-Bicocca, Piazza della Scienza 3, 20126 Milano, Italy}
\address{$^2$ INFN, Sezione di Milano-Bicocca, Piazza della Scienza 3, 20126 Milano, Italy}
\address{
    $^3$ Departament de F\'isica, Universitat de les Illes Balears, IAC3-IEEC,\\
    Ctra.  Valldemossa km 7.5, E-07122 Palma, Spain
}
\address{
    $^4$ University of Portsmouth, Portsmouth, PO1 3FX, United Kingdom
}
\address{
    $^5$ SUPA, School of Physics and Astronomy, University of Glasgow,\\
    Glasgow G12 8QQ, United Kingdom
}
\address{
    $^6$ Kaggle/Google, Mountain View CA, USA
}
\address{$^7$ Independent researcher}
\address{$^8$ Vergesense, Mountain View 94041 CA, USA }
\address{$^9$ University of Virginia, Department of Materials Science and Engineering, Charlottesville, VA 22904-4745, USA}
\address{$^{10}$ Aillis Inc., Tokyo, Japan}
\address{$^{11}$ Department of Health Policy and Public Health, Graduate School of Pharmaceutical Sciences, The University of Tokyo, Tokyo, Japan}
\address{$^{12}$ CS Group, Tolouse, 31500, France}
\address{$^{13}$ Departamento de Física Teórica and Instituto de Física Teórica UAM/CSIC, Universidad Autónoma de Madrid, Cantoblanco, 28049, Madrid, Spain}
\address{$^{14}$ Department of Computer Science, University of Copenhagen, Universitetsparken 1, DK-2100,  Copenhagen, Denmark}
\address{$^{15}$ Niels Bohr Institute, University of Copenhagen, Jagtvej~155A, DK-2200, Copenhagen, Denmark}

\ead{rodrigo.tenorio@unimib.it}

\date{\today}

\begin{abstract}
    We report results of a public data-analysis challenge,
    hosted on the open data-science platform Kaggle,
    to detect simulated continuous gravitational-wave signals (CWs).
    These are weak signals from rapidly spinning neutron stars that remain 
    undetected despite extensive searches.
    The competition dataset consisted of a
    population of CW signals using both simulated and real LIGO detector 
    data matching the conditions of actual CW searches. 
    The competition attracted more than 1,000 participants to develop
    realistic CW search algorithms. 
    We describe the top 10 approaches and discuss their applicability as a
    pre-processing step compared to standard CW-search approaches.
    For the competition's dataset, we find that top approaches can reduce the 
    computing cost by 1 to 3 orders of magnitude at a false-dismissal probability 
    comparable to standard CW searches. 
    Additionally, the competition drove the development of new 
    GPU-accelerated detection pipelines, which facilitated 
    their adoption in other areas of gravitational-wave data analysis.
    We release the associated dataset, which constitutes the first open 
    standardized  benchmark for CW detection, to enable reproducible method comparisons 
    and to encourage further developments toward the first detection of 
    these elusive signals.
\end{abstract}

\maketitle

\section{Introduction\label{sec:introduction}}

Continuous gravitational waves (CWs) are long-duration gravitational-wave (GW) 
signals expected from rapidly-spinning  non-axisymmetric neutron stars (NSs)~\cite{Riles:2022wwz}.
Although currently undetected,their detection would shed light on the extreme physics of NSs~\cite{Haskell:2023yrv}, 
allow us to understand their population in our galaxy~\cite{Pagliaro:2023bvi, Hua:2023aff},
or even establish a new probe to direct and indirect detection 
of dark matter~\cite{Miller:2025yyx}.

CW searches can be classified according to their assumptions on the expected source~\cite{Wette:2023dom}.
Targeted searches, for example, assume the CW emission of a pulsar is phase-locked to its electromagnetic emission;
this allows for very sensitive searches using matched filtering.
Blind searches, on the other hand, attempt to detect a CW signal from unknown sources;
due to the sheer size of their parameter-space, these kinds of searches cannot make use of matched filtering.

A standard CW search operates by filtering data against a set of signal models,
usually referred to as \emph{templates}, which are parameterised by their amplitude and 
phase-evolution parameters (see Sec.~\ref{subsec:signals}). 
The result of filtering data against a specific signal
model is a \emph{detection statistic}, which is then used to establish the significance of
a candidate. The specific detection statistic in use depends on search assumptions and 
available computing budget, and has a direct impact on sensitivity~\cite{Tenorio:2021wmz}.

Fully-coherent searches typically use the \mbox{$\mathcal{F}$}-statistic~\cite{Jaranowski:1998qm,Cutler:2005hc},
which tracks the phase of a CW signal accounting for possible amplitude and frequency modulations due
to the orbital motion of GW detectors. This approach is unfeasible for broad parameter spaces as the number
of required signal templates quickly becomes computationally prohibitive~\cite{Wette:2015lfa}.
Semicoherent detection statistics, on the other hand, follow the frequency evolution of a signal and
only track its phase evolution along short segments. Effectively, a semiocherent statistic accumulates 
coherent statistics (such as segment-wise $\mathcal{F}$-statistics, power spectra~\cite{soap, Tenorio:2024jgc}, 
or the binarized power~\cite{Krishnan:2004sv, 2014PhRvD..90d2002A}) following the signal's frequency evolution.
These statistics impose milder constraints and require less templates to cover a given parameter-space region, 
reducing the computing cost of a search, but also reduce the sensitivity as background  
noise is more likely to fit a signal template~\cite{Prix:2012yu}. 
Because of this, semicoherent searches are robust to unmodelled physics such as
NS glitches~\cite{Ashton:2017wui} or spin-wandering~\cite{Mukherjee:2017qme, Carlin:2025sxm}.

The current trend in blind CW searches is to explore relatively narrow parameter-space
regions to aim for astrophysically plausible CW sources~\cite{Dergachev:2025ead,Dergachev:2025hwp,
Steltner:2023cfk,KAGRA:2022dwb}, or to extend the parameter space towards binary 
systems~\cite{LIGOScientific:2020qhb,Covas:2022rfg,Covas:2024nzs}, which may emit detectable CW
signals more easily due to accretion~\cite{Ushomirsky:2000ax, Haskell:2015psa, Hutchins:2022chj}.
The most sensitive of these searches are based on highly-efficient implementations of the
semicoherent $\mathcal{F}$-statistic, 
such as the Global Correlation Transform (\eah{})~\cite{Pletsch:2009uu},
\weave{}~\cite{Wette:2018bhc}, or \texttt{BinarySkyHou}$\mathcal{F}$~\cite{Covas:2022mqo}.
These searches require significant computing budgets to be successfully deployed.
Alternatively, short-coherence 
searches followed by hierarchical follow-ups~\cite{Ashton:2018ure,
Tenorio:2021njf,Covas:2024pam,Mirasola:2024lcq},
or signal-agnostic searches such as \texttt{SOAP}~\cite{soap},
require a much more modest budget, albeit with a corresponding reduction in sensitivity.
All in all, the sensitivity of a CW search is primarily limited by the available computing 
resources~\cite{Prix:2012yu}.

Machine learning applications~\cite{Cuoco:2024cdk} have demonstrated significant accelerations in 
search~\cite{Gabbard:2017lja,Chua:2019wwt,Koloniari:2024kww,Nagarajan:2025hws}
and parameter-estimation~\cite{Gabbard:2019rde,Dax:2021tsq,Williams:2021qyt,Williams:2023ppp} 
workflows targeting compact binary coalescences compared to standard  
approaches~\cite{Usman:2015kfa,Sachdev:2019vvd,Cornish:2020dwh}.
Due to the comparatively broader parameter space, applications to blind CW signals have reported
limited success~\cite{Dreissigacker:2019edy,Yamamoto:2020pus,Joshi:2025xdz,Cheung:2025lyf},
with most successful applications focusing on accelerating specific
computational steps~\cite{Modafferi:2023nzt} or increasing the robustness of a search to 
unmodelled physics~\cite{Bayley:2020zfa,Bayley:2022hkz}.

In this work, we present the results of an open data-analysis competition
to spark new approaches and algorithmic developments in CW searches~\cite{competition}.

We constructed a dataset using the sensitivity of flagship searches such as \texttt{GCT} or \texttt{Weave}
as a baseline reference to maximise the impact of winning solutions, 
and compared their performance to an agnostic CW search (\texttt{SOAP}) to gauge the impact
of design choices made by different teams.
This competition delivered a comparison of methods beyond what within-domain 
mock data challenges~\cite{Messenger:2015tga,Walsh:2016hyc,
Schafer:2022dxv,Baghi:2022ucj} are able to prospect and allowed us to understand the complications and
computational advantages achievable by off-the-shelf approaches from beyond the community.
We release the dataset of the competition~\cite{dataset} to provide the first
publicly-available benchmark targeted towards the detection of CW searches.
Tools and tutorials to interact with this data are publicly available
in Kaggle~\cite{kaggle_post} and the~\texttt{PyFstat} repository~\cite{pyfstat_github}.

The paper is structured as follows: In Sec.~\ref{sec:leveraging} we introduce the rationale
behind this CW-search data-analysis competition.
In Sec.~\ref{sec:dataset} we introduce the competition's dataset.
In Sec.~\ref{sec:winning_solutions} we discuss the top-scoring
solutions of the competition. In Sec.~\ref{sec:cw_interpretation} we interpret these solutions 
in the context of standard CW searches. We finish by discussing these results
in Sec.~\ref{sec:discussion}. We also summarize the results of the first Kaggle competition 
aimed at detecting GW signals from compact binary coalescences~\cite{first_challenge} 
in~\ref{sec:first_kaggle}.

\section{Leveraging open competitions for gravitational-wave research\label{sec:leveraging}}

The development of CW search methods is often constrained by the expertise and computational
perspective available within the CW research community. The complexity of detecting a signal
in noisy data, however, represents a challenge that may benefit, if properly supervised and tested,
from a fresh perspective as provided by the broader data science community.

Open data-analysis competitions have demonstrated significant value in advancing 
scientific research across multiple domains, such as medical 
imaging~\cite{lungCompetition,BrainTumorPrediction,RNAprediction},
time-series forecasting~\cite{forecasting}, or particle physics~\cite{Calafiura:2018cdd}, 
often producing solutions that outperform existing standards.
These successes stem from competitions' ability to attract diverse 
problem-solving approaches, encourage rapid iteration,
and provide standardized benchmarks for method comparison~\cite{CompetitionsAreGold}.

The gravitational-wave community has already embraced crowd-sourcing approaches through platforms,
such as  Zooniverse~\cite{zooniverse}, 
to understand transient noise in interferometric detectors~\cite{Zevin:2016qwy,Zevin:2023rmt,Razzano:2022lgg},
or Kaggle~\cite{kaggle} to search for gravitational-wave signals from compact binary
coalescences~\cite{first_challenge}. 
This approach can be understood as an evolution of the well-established mock data challenges
conducted by the GW community~\cite{Messenger:2015tga,Walsh:2016hyc, Schafer:2022dxv,Baghi:2022ucj},
where the aim has shifted from comparing existing pipelines to exploring new strategies altogether.
The overall experience in GW citizen-science so far suggests non-experts,
when given a proper training framework, can deliver solutions beneficial 
to the GW community~\cite{Soni:2021cjy,Mackenzie:2025gut}. 
For example, the results of the first Kaggle competition~\cite{first_challenge} served as a basis
to push forward our understanding of GW searches using machine learning~\cite{Nagarajan:2025hws}.

The problem of detecting CW signals presents an ideal case for an open competition: 
First, the problem can be framed as a well-defined classification 
task~\cite{Prix:2009tq, Searle:2008jv}, which provides a simple criterion
to rank different strategies against each other. Second, synthetic data can be easily
generated, either from first principles or through well-established software
packages~\cite{lalsuite,swiglal,pyfstat}. Third, since large-scale searches can be understood
as the aggregation of many small-scale searches~\cite{Tenorio:2024jgc}, competitors can tackle
a simulation of the real problem using standard hardware. Finally, despite the existing experience
in the field, large portions of the search-pipeline configuration space where novel improvements may
be found remain yet to be explored.

Building on previous positive results~\cite{first_challenge}, we released our competition on 
Kaggle~\cite{kaggle}, a well-established data-analysis platform powered by Google with proven track record 
in scientific applications and more than a million active users spread across vast domains of 
data science and engineering.

\section{Dataset description\label{sec:dataset}}

We seek a data-analysis strategy compatible with a blind CW search; that is, a pipeline
which classifies narrow frequency bands of LIGO/Virgo/KAGRA~\cite{LIGOScientific:2014pky,
VIRGO:2014yos,KAGRA:2018plz,KAGRA:2020tym} data according to whether they
contain a signal or not. We take the sensitivity of deep searches conducted 
using \eah{}~\cite{eah_O2}  and \weave{}~\cite{Wette:2021tbv} as a reference
for our competition. These pipelines use highly-efficient implementations of the semicoherent 
$\F$-statistic~\cite{Pletsch:2009uu, Wette:2018bhc, Dunn:2022gai}
to deliver the most sensitive all-sky search results across their explored parameter space.
Their sensitivity is effectively bounded by the available computing resources, 
which amount to about 100 million CPU hours or 1 million GPU hours, depending on the specific
implementation. By framing CW detection as a binary classification problem, 
we aim to discover whether alternative approaches could compete with or complement 
these well-established strategies. Note that the high computing cost of these searches
complicates a direct comparison of methods; instead, we use the sensitivity of these
pipelines to construct a signal distribution for the challenge, and after the competition
discuss the advantages of the methods presented by the competitors.

\subsection{Data format}

We constructed a dataset using Short Fourier Transforms~\cite{SFTFormat} 
(SFTs), a data format used by actual blind CW  searches. 
This choice provides the participants with the same information
available to CW data analysts and simplifies the transition of a successful solution into a production stage.

Given a set of $N$ time-domain data samples $x_{\alpha}[j],  j=0, \dots, N-1$ 
starting at an epoch $t_{\alpha}$ sampled with a period $\Delta t$, we define an SFT following~\cite{SFTFormat}: 
\begin{equation}
    \tilde{x}_{\alpha}[k] 
    = \Delta t \sum_{j=0}^{N-1} x_{\alpha}[j] e^{-\frac{-2 i \pi j k}{N}} \,.
\end{equation}
Here the index $k=0, \dots, N/2$ labels the positive frequencies $f_k = k / \T{SFT}$, where
$\T{SFT} = N \Delta t$ is the duration of an SFT. 
These frequencies physically correspond to $F_0 + f_k$, where the fundamental frequency $F_0$
is always extracted through a homodyne filter before computing the SFT.

The competition dataset consisted of 8,000 test samples divided into signal and noise samples
in equal parts. Additionally, 600 training samples were provided to the competitors as a quick
entrypoint to the competition. The dataset was made available on the Kaggle platform, and adds up
to $\sim\!0.5\, \mathrm{TB}$. 
Additionally, we supplied the tools to generate simulated CW data through
the \texttt{PyFstat} package~\cite{pyfstat}, which provides a Python interface to the
LIGO/Virgo/KAGRA  Algorithm Library~\cite{lalsuite, swiglal}.

The dataset contains both noise-only samples and software-simulated CW signals, 
using both simulated Gaussian noise and data slices from the third observing run of the 
Advanced LIGO detectors~\cite{KAGRA:2023pio, O3aData, O3bData}. 
The chosen frequency band ($\SI{0.2}{\hertz}$ or $\SI{0.2}{\hertz} \times \T{SFT} = 360$
frequency bins) is compatible with those analyzed by current blind searches. The sample duration,
on the other hand, falls on the shorter end, as most searches analyse at least 6 months of data. 
This choice is due to the limited storage available on the Kaggle platform. Regardless, the 
duration is long enough to contain the main distinctive features of CW signals, namely a
Doppler shift on the time-scale of months combined with a daily amplitude modulation.

\subsection{Noise distribution\label{sec:noise}}

All samples of the competition contain background noise. This was either simulated
Gaussian noise or real Advanced LIGO noise from the third LVK observing run 
(O3)~\cite{KAGRA:2023pio, O3aData, O3bData}. 

Each data sample in the competition's dataset consists of a two sets of 
Short Fourier Transforms, representing the H1 (Hanford) and L1 (Livingston) 
LIGO detectors,  spanning a total time of $\T{span} = 4\;\mathrm{month}$ and 
a frequency band of $\SI{0.2}{\hertz}$ with $\T{SFT} = \SI{1800}{\second}$. 
To better simulate the conditions of a real search, we introduce gaps in the data.
As a result, SFTs are not necessarily contiguous  (i.~e.~\mbox{$t_{\alpha + 1} - t_{\alpha} \geq \T{SFT}$}).

\begin{figure}
    \centering
    \includegraphics[width=0.6\textwidth]{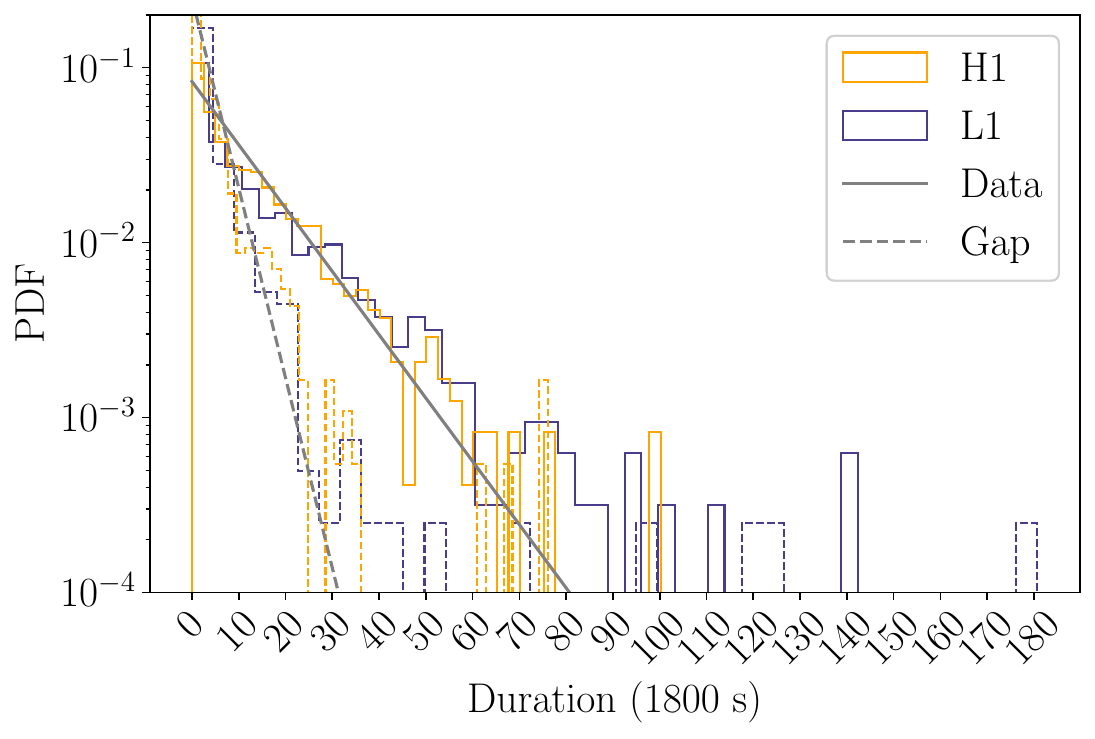}
\caption{
    Distribution of observing-mode (solid line) and gap (dashed line)
    duration throughout the O3 observing run. Different colors refer
    to different detectors. Gray lines are an exponential fit as discussed
    in the text.
}
\label{fig:data_segments}
\end{figure}

We model the duration of observing-quality (``science-mode'') segments and gaps
during the O3 run using an exponential distribution, as shown in Fig.~\ref{fig:data_segments}.
We compute the duration of science-mode segments and gaps by generating 
\SI{1800}{\second}-SFTs from publicly-available O3 data using the timestamps provided in Ref.~\cite{segments}. 
The average science-mode segment (gap) duration is compatible with
\mbox{$\chtime{data} = \qtyproduct{2.16e4}{\second}$}
(\mbox{$\chtime{gap} = \qtyproduct{7.2e3}{\second}$}),
which corresponds to $\sim$12 ($\sim$4) SFTs. 

For each interferometer, we draw gap/science-mode durations alternatively 
from the distributions in Fig.~\ref{fig:data_segments} until completing a span of $\T{span}$.
The first segment is sampled from the gap distribution and added to the starting
time of O3 to avoid fixing the starting time. 
As a result, the SFTs of different detectors have, in general, different starting timestamps.
By construction, the resulting data samples for both detectors
have a ``duty cycle'' (fraction of science-mode data) of
\begin{equation}
    \dc = \frac{\chtime{data}}{\chtime{data} + \chtime{gap}} \simeq 0.78 \,,
\end{equation}
which is comparable with realistic observing runs~\cite{Buikema_2020,LIGO:2021ppb}.

Gaussian-noise data sample are generate by drawing from a Gaussian distribution with zero mean and
(single-sided) Amplitude Spectral Density (ASD) \mbox{$\sqrt{\Sn} = \qtyproduct{5e-24}{\persqrthz}$}.
Equivalently, the real and imaginary parts of each complex SFT bin $\tilde{x}_{\alpha}[k]$ are drawn
from a Gaussian distribution with zero mean and standard deviation $\frac{1}{2} \sqrt{\T{SFT} \Sn}$.
This setup is comparable with mock analyses conducted in current CW searches 
(see e.g.~\cite{Tenorio:2024jgc} and references therein).

To generate a data sample containing real noise, we instead sample SFTs from LVK O3 data.
To prevent participants from identifying noise realisations using publicly available data,
we shift the frequency labels by a random amount within \SI{\pm 4}{\hertz} and add Gaussian 
noise with 0.1\% of the data's ASD to mangle the numerical noise values but maintain real-noise features.
This value was chosen empirically.

\subsection{Signal distribution\label{subsec:signals}}

CW signals from isolated NSs are typically described in terms of 
the amplitude parameters $\A$, comprising the GW's nominal amplitude $h_0$,
the (cosine of the) inclination angle with respect to the line of sight $\cos\iota$,
the polarisation angle $\psi$, and the initial phase $\phi_0$; and the phase-evolution
parameters $\lambda$, which include the signal's frequency $f_0$ and frequency derivative 
(spindown) $f_1$ at a fiducial reference time $t_{\mathrm{ref}}$, 
and the sky location $\hat{n}$. We take the reference time at the start of the O3 run and
parameterise the sky position using equatorial coordinates (right ascension
$\alpha$ and declination $\delta$). 

We simulate a source population compatible with that assumed in blind searches,
as summarized in Table~\ref{table:signal_distributions}. To generate a signal sample, 
we draw a set of parameters $(\A, \lambda)$ from the population, we simulate noise as
described in Sec.~\ref{sec:noise} for both the H1 and L1 LIGO detectors, 
and use \texttt{MakeFakeData\_v5}~\cite{lalsuite} to add a simulated signal 
into the noise. This software simulates a CW signal as observed by a ground-based
interferometric detector, adds Gaussian noise with a given ASD, and returns a the
corresponding SFTs with the specified $\T{SFT}$.

The sky position $\hat{n}$ is drawn uniformly across the celestial sphere; 
the distribution over the initial phase is uniform along $[0, 2 \pi)$;
orientation angles $(\cos \iota, \psi)$ are drawn following an isotropically-oriented population. 
The spindown $f_1$ is drawn  from a log-uniform distribution along 
$[-10^{-8}, -10^{-12}] \, \mathrm{Hz}/\mathrm{s}$ with 90\% probability 
and $[10^{-12}, 10^{-9}]\, \mathrm{Hz}/\mathrm{s}$ with a 10\% probability.
This split accounts for possible spin-ups due to accretion, proper motion,
or the evaporation of a boson cloud~\cite{Jaranowski:1998qm, Mukherjee:2017qme,Zhu:2020tht} 

The frequency $f_0$ is sampled uniformly along $[50, 500]\,\mathrm{Hz}$,
which is a frequency range commonly explored by blind CW searches as it 
contains the most sensitive band of the Advanced LIGO detectors~\cite{Buikema_2020}.
The bandwidth of a CW signal depends on 
its frequency and sky location, as it is largely dominated by the Doppler effect
due to the orbital motion of the detector around the Sun. For the parameter space
considered in this challenge, the maximum bandwidth is about \SI{0.05}{\hertz}. 
To complete a band of \SI{0.2}{\hertz}, we attach noise data to both sides of the 
signal's spectrum so it is contained completely within the data sample.
Also, to avoid biasing the dataset, the proportion of data attached to each side
of the signal's frequency band is chosen by drawing a uniform random number between 0 and 1
so that the signal is not systematically centred. 

\begin{table}
\centering
\begin{tabular}{rccc}
    \toprule
    Parameter & Unit & Symbol &  Distribution \\
    \midrule
    Frequency & \unit{\hertz} &  $f_0$ &  $\mathrm{Uniform}(50, 500)$\\
    Spindown & \unit{\hertz\tothe{2}} & $f_1$ & 
    $-10^{-\mathrm{Uniform}(8, 12)} \, (90\%)$\\
    &&& $+10^{-\mathrm{Uniform}(9, 12)} \, (10\%) $\\
    Right Ascension &  \unit{\rad} & $\alpha$ & $\mathrm{Uniform}(0, 2\pi)$\\
    (sin) Declination & -- & $\sin{\delta}$ &  $\mathrm{Uniform}(-1, 1)$\\
    (cos) Inclination angle &  -- & $\cos{\iota}$ & $\textrm{Uniform}(-1, 1)$\\
    Polarization angle & \unit{\rad} & $\psi$ & $\textrm{Uniform}\left(-\pi/4, \pi/4\right)$\\
    Initial phase & \unit{\rad} & $\phi_0$ & $\textrm{Uniform}\left(0, 2\pi\right)$\\
    SNR & -- & $\rho$ & $\Gamma(5, 9)$\\
    \bottomrule
\end{tabular}
    \caption{
        Distribution of signal parameters used to generate the competition's challenge.
        The Uniform distributions are defined by their lower and upper limits;
        the gamma distribution $(\Gamma)$ is expressed using the shape-scale parameterisation.
    }
    \label{table:signal_distributions}
\end{table}

We describe the amplitude of a CW signal using the Signal-to-Noise ratio (SNR)
\begin{equation}
    \rho(h_0, \psi, \cos\iota, \hat{n})
    = \sqrt{\frac{\T{data}}{\Sn}} g(\psi, \cos\iota, \hat{n}) h_0 
\end{equation}
where $g(\psi, \cos\iota, \hat{n})$ is a known 
function~\cite{Dreissigacker:2018afk} encoding the dependency on the 
orientation angles and sky position, and $\T{data} = \T{SFT} N_{\mathrm{SFT}}$,
where $N_{\mathrm{SFT}}$ refers to the \emph{total} number of SFTs (counting each detector separately).
The SNR is directly related to the detectability of a 
signal~\cite{Jaranowski:1998qm}, and is proportional to $h_0$ given
$(\psi, \cos\iota, \hat{n})$ and a dataset. This implies the SNR distribution will 
determine the difficulty of the challenge, as high SNR signals
($\rho  > 80$ in our specific configuration) tend to display recognizable features
in the data, making them easy to detect. 

Blind CW searches usually quote their results in terms of
population-based upper limits at a certain false-dismissal probability 
(usually 5\% to 10\%) on a constant-$h_0$ 
population~\cite{Tenorio:2021wmz,Wette:2023dom}.
The false-alarm probability of a search is usually not computed.
This complicates the comparison of Kaggle solutions to realistic searches
on equal footing. Instead, we will base our comparisons on computing
cost; concretely, we will assess the capability of Kaggle solution to discard
broad parameter-space regions in such a way that signals detectable by a blind
CW search are not discarded. To do so, it will suffice to re-scale the sensitivity
estimates reported by the two reference searches to the dataset duration used
in the competition.

The two reference searches we use in this challenge, based on the
\eah{}~\cite{eah_O2} and \weave{}~\cite{Wette:2021tbv} pipelines, 
report their sensitivity estimates in terms of the average amplitude 
$\langle h_0\rangle$ at which 90\% of an isotropically-oriented 
uniformly-sky-distributed signals would be detected.
We relate this quantity to the corresponding
averaged SNR as~\cite{Wette:2011eu}
\begin{equation}
    \sqrt{\snrSqAvASky} = \frac{2}{5} \sqrt{\frac{\T{data}}{\Sn}} \langle h_0 \rangle \;.
    \label{eq:averaged_optimal_snr}
\end{equation}
These pipelines split the dataset into segments with a duration $\T{coh}$
which are analysed independently using the $\F$-statistic,
and then the results are combined to compute the semicoherent
$\F$-statistic.

We re-scale the sensitivity estimates using $\T{data}$, 
which corresponds to increasing $\T{coh}$
so that the number of coherent segments is unchanged. 
The original searches used O2  Advanced LIGO data, 
with a nominal duration of 9 months and about 
\SIrange{60}{65}{\percent} duty cycle~\cite{LIGOScientific:2019lzm}: 
$\T{data}^{\mathrm{O2}} \approx \SI{2.6e7}{\second}$.
Our challenge covers a nominal duration of 4 months and a duty cycle of 
$78\%$: $\T{data}^{\mathrm{K}} \approx \SI{1.6e7}{\second}$.
The sensitivity of these searches in the competition's dataset corresponds
to scaling  their reported sensitivity up by 
$\sqrt{\T{data}^{\mathrm{O2}} /\T{data}^{\mathrm{K}}} \approx 1.3$,
as shown in Table~\ref{table:snrs}.

\begin{table}
    \centering
    \begin{tabular}{rcccc}
        \toprule
        & $\T{coh}\;(\unit{\hour})$ & $\avg{h_0}$ 
        & $\sqrt{\snrSqAvASky}$ & $\sqrt{\snrSqAvASky^{\mathrm{K}}}$ \\
        \midrule
        \eah{}~\cite{eah_O2} & 60 & $1.3 \times 10^{-25}$ &  47 & 61 \\
        \weave{}~\cite{Wette:2021tbv} & 240 &  $1.0 \times 10^{-25}$  & 36 & 47\\
        \bottomrule
    \end{tabular}
    \caption{
        Summary of sensitivity estimates reported by the two searches
        considered in this work. These are computed by averaging the reported
        sensitivity estimates across the frequency band of the search.
    }
    \label{table:snrs}
\end{table}

\begin{figure}
    \centering
    \includegraphics[width=0.6\textwidth]{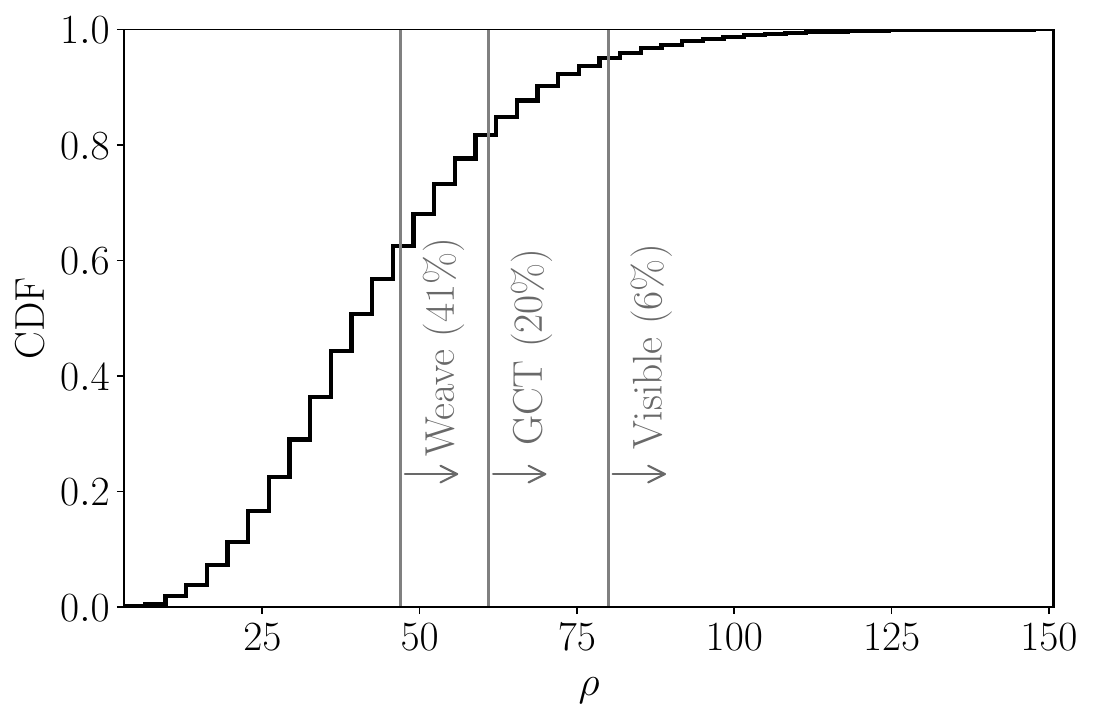}
    \caption{
        Empirical cumulative distribution of SNR $\rho \sim \Gamma(5, 9)$. 
        Vertical lines denote the re-scaled sensitivity of different search 
        pipelines as discussed in the main text. 
    }
    \label{fig:snr_distro}
\end{figure}

\begin{figure}
    \centering
    \includegraphics[width=\linewidth]{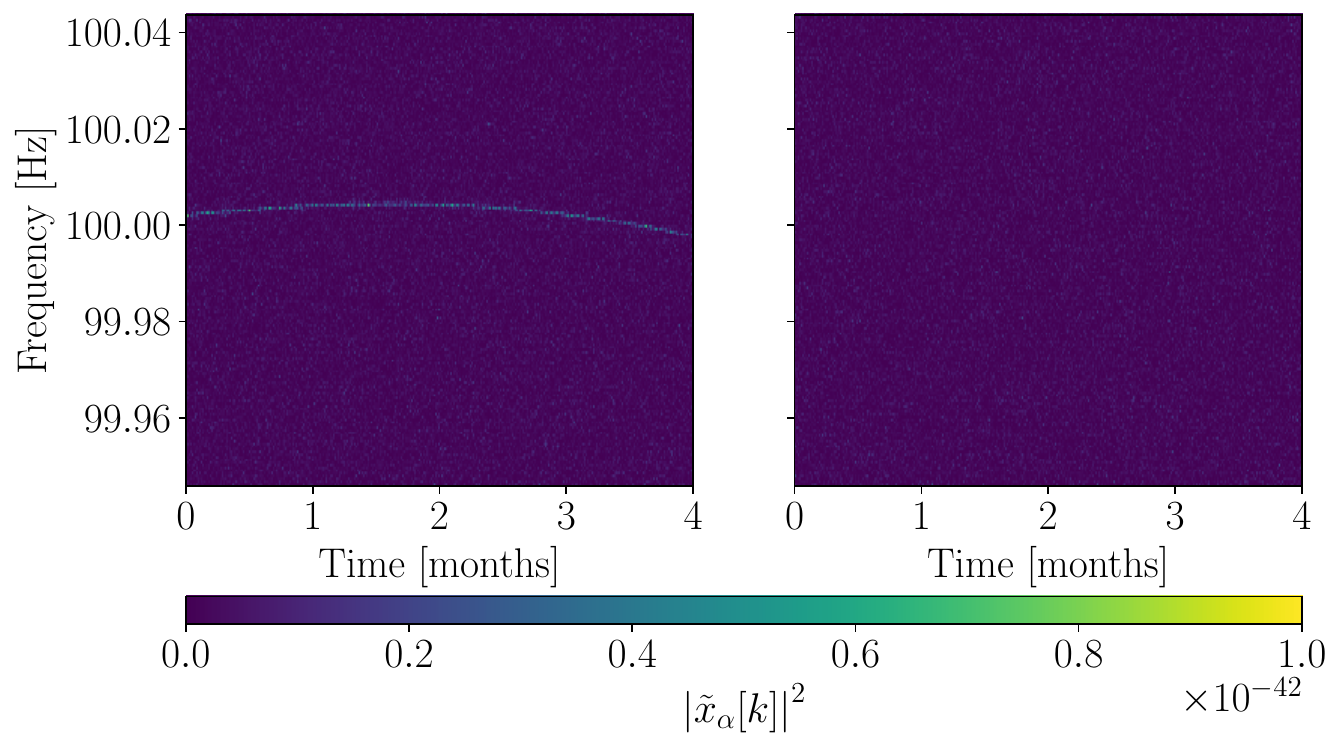}
    \caption{
    Spectrograms of a sample containing 4 months of simulated Gaussian noise
    with a fiducial CW signal as observed by the H1 detector.
    The CW amplitude in the left (right) spectrogram is such that $\rho = 350$ ($\rho = 35$);
    the signal parameters are otherwise identical. As a result, the signal is visible in the
    left spectrogram but not in the right one. This data was generated without gaps for the sake
    of visualization.}
    \label{fig:example_spectrogram}
\end{figure}

With this information, we choose to describe the SNR distribution
using a Gamma distribution 
with shape parameter $k=5$ and scale parameter $\theta = 9$,
shown in Fig.~\ref{fig:snr_distro}.
This distribution is such that 20\% of the signals are above the sensitivity 
of \eah{} and about 41\% of the signals are above the sensitivity of \weave{}. 
6\% of the signals display visible features ($\rho > 80$). 
We show an example CW signal at different SNR levels in Fig.~\ref{fig:example_spectrogram}.

\section{Competition results\label{sec:winning_solutions}}

The Kaggle competition was held from October 4th 2022  to January 4th 2023,
and accumulated a total number of  $\sim$30,000 submissions amongst 936 teams.
Competitors were provided with a dataset as described in Sec.~\ref{sec:dataset},
and the task was to submit a ranking statistic for each sample in order to classify
them as signal or noise. 

Each submission was ranked according to the Area Under the Receiver Operating 
Characteristic (ROC) curve (AUC score).  That is, the ranking statistics of a 
submission where used as thresholds $t$ to compute the false-dismissal 
probability (fraction of signal samples classified as noise, $p_{\mathrm{FD}}(t)$)
at a given level of false-positive probability (fraction of noise samples classified
as signals, $p_{\mathrm{FP}}(t)$). The AUC was then computed by integrating the area
under the curve described by $(p_{\mathrm{FP}}(t), p_{\mathrm{FD}}(t))$.
This ranking metric was the most suitable for our data-analysis problem given the
available selection in the Kaggle platform. 

The challenge actually makes use of two leaderboards, namely \emph{public} 
leaderboard, which is visible throughout the duration of the competition,
using a small fraction of the competition's dataset to compute
the AUC score, and the \emph{private} leaderboard, which computes the AUC
score using the complete dataset. This design allows participants to obtain
a continuous assessment of their performance against other competitors,
which is beneficial for a rapid iteration of the solution's design, 
and minimises the chances of winning solutions ovefitting to a specific dataset.
Additionally, the use of a random subset of the dataset minimises the probability
of competitors successfully inferring the true classification of test data samples
through repeated sumbmission~\cite{Whitehill15,Whitehill17climb,Whitehill17}.

\begin{figure}
    \includegraphics[width=\textwidth]{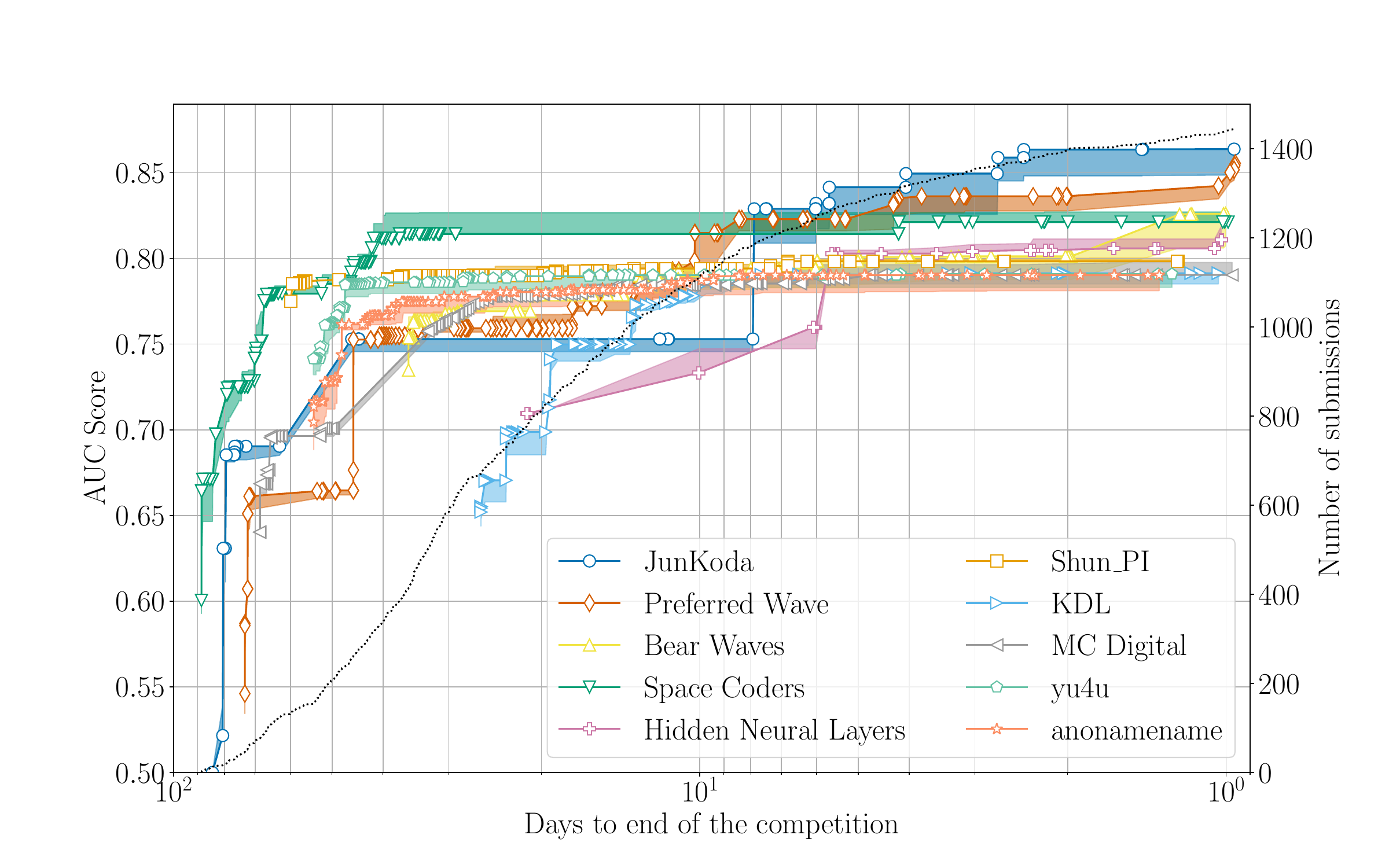}
    \caption{
        Evolution of the top-10 teams throught the competition in terms 
        the AUC score. The markers correspond to the private leaderboard score, 
        while the shaded regions extend to the public leaderboard score.
        The cumulative number of submissions by these 10 teams is shown as a
        dotted line.
    }
    \label{fig:score_evolution}
\end{figure}

We show in Fig.~\ref{fig:score_evolution} the evolution of the leaderboard
scores for the top 10 teams of the competition. 
Overall, one can distinguish two periods in the competition: From  the start of
the competition to about 10 days before the end the leaderboard was 
dominated by Space Coders. Ten days before the end, three more teams
(\team{JunKoda}, \team{PreferredWave}, and \team{BearWaves}) managed to surpass 
their score. Overall, most teams converged rapidly into a sensible
solution whose sensitivity remained broadly unchanged throughout the challenge.

\subsection{Top-scoring submissions}

We divide the solutions proposed by the top-10 scoring teams into three broad classes: 
Track-Statistic, Track-Fitting, and Machine-Learning.

\subsubsection{Track-Statistic\label{sec:track_stat}} 

The first class contains solutions implementing a track-based detection statistic
similar to those used in blind CW searches with short coherence times.
Specifically, these methods assume a closed-form frequency evolution 
(i.e. a ``frequency track'' in the spectrogram) parameterised by a small
number of parameters  (such as frequency, spindown, and sky position) along which a detection
statistic is accumulated. The search then consists in finding the parameters that maximise
said detection statistic. This class includes the teams \team{JunKoda}~\cite{first_team}~(1st),
\team{PreferredWave}~\cite{second_team}~(2nd), \team{HiddenNeuralLayers}~\cite{fifth_team}~(5th),
and \team{Shun\_PI}~\cite{sixth_team}~(6th).
These methods present two main degrees of freedom, namely the detection statistic and strategy to cover the parameter space, which we summarize in
Table~\ref{table:track_stat}. To address non-Gaussianities arising in 
real-data samples, all teams masked frequency bands in the spectrogram 
containing large deviations from Gaussian noise.

\begin{table}
    \scriptsize
    \centering
    \begin{tabular}{lcccc}
        \toprule
        & GPU & Statistic & Template placement \\
        \midrule
        \team{JunKoda} (1st)
            & Yes
            & Maximum Dirichlet-weighed power~\cite{Allen:2002bp}
            & Square grid ($3.5 \times 10^{8}$ templates) \\
        \team{PreferredWave} (2nd)
            & Yes
            & Average power
            & Uniform sampling ($3.3 \times 10^{6}$ templates) \\
        \team{HiddenNeuralLayers} (5th)
            & No
            & Maximum power
            & Differential evolution (100 steps)~\cite{DiffEv}\\
        \team{Shun\_PI} (6th)
            & No
            & Maximum power
            & Simulated-annealing~\cite{SimAn}. \\
        \bottomrule
    \end{tabular}
    \caption{Design choices of the winning solutions presented by the teams
    in the Track-Statistic class of solutions.}
    \label{table:track_stat}
\end{table}

The relative placement of these four teams in the leaderboard is consistent with
their strategies. \team{HiddenNeuralLayers} and \team{Shun\_PI} place templates in a 
data driven manner. While such a strategy has been proven effective in local 
analyses~\cite{Ashton:2017wui, Tenorio:2021njf, Covas:2024pam, Mirasola:2024lcq}, 
the breadth of the CW parameter-space makes them unsuitable for broad searches 
such as the ones here considered~\cite{Ashton:2017wui},
hence their relatively lower placement. The difference between \team{JunKoda} and 
\team{PreferredWave}, on the other hand, can be ascribed to the
significant difference in the number of evaluated templates ($10^8$ in 
\team{JunKoda} versus $10^{6}$ in $\team{PreferredWave}$) and the fact that 
\team{JunKoda} takes the maximum rather than the average power in the
template bank~\cite{Tenorio:2024jgc}.

\subsubsection{Track-Finding}

The second class introduces an extra degree of freedom with respect to Sec.~\ref{sec:track_stat},
namely that the frequency evolution of the signal is not parameterised but inferred
from the data. This approach is used by team \team{SpaceCoders}~(4th),
and bears some similarities to CW searches based on the Viterbi algorithm~\cite{soap}.

Specifically, the method uses dynamic programming to find a track maximising the accumulated
SFT power. The track is constrained to contain a single inflection point, given the duration
of the dataset and the expected signal morphology. Upon identifying a track, the signal's parameters
are estimated using a parametric fit in terms of frequency, spindown, and sky position. The optimisation
is carried out using the Nelder-Mead algorithm~\cite{nelder_mead}, in a similar manner to the methods
presented in Sec.~\ref{sec:track_stat}.

The relative loss of \team{SpaceCoders} with respect to other track-based methods can be justified as follows:
First, the number of tracks explored by the \team{SpaceCoders} search is exponentially greater than a
closed-form track search~\cite{Wette:2023dom}; this shifts the expected 
background distribution upwards~\cite{Tenorio:2021wad} and, consequently, lowers the detection probability
at a fixed amplitude. Second, the track partially relies on the signal displaying strong features within
short timescales to clearly identify it. Due to the idiosyncrasy of CWs, this is only true for relatively 
strong signals.

\subsubsection{Machine-Learning}

The final class includes solutions using machine-learning strategies; that is, search pipelines in which
some component is \emph{trained} before producing a submission. This class comprises teams 
\team{BearWaves}~\cite{third_team}~(3rd), \team{KDL}\footnote{No available information about this team}~(7th), \team{MCDigital}~\cite{eighth_team}~(8th), 
\team{yu4u}~\cite{nineth_team}~(9th), and \team{anonamename}~\cite{tenth_team}~(10th). 

\team{anonamename} implemented a classifier based on a Convolutional Neural Network (CNN).
Data samples are pre-processed following~\cite{tenth_team} and two outputs are produced:
a detection statistic, and the estimated location of the signal in the spectrogram as a mask. 
To account for real data, line-like features were generated and included during training.

\team{yu4u}~\cite{nineth_team} implemented a similar strategy using a U-net architecture to estimate 
the location of the signal in the spectrogram and compute a detection statistic. 
As in the previous case, real data noise features were addressed by generating data containing line-like features
during training.

\team{MC Digital}~\cite{eighth_team} combines two detection statistics using a CNN and the $\F$-statistic
as computed by \texttt{PyFstat}~\cite{pyfstat}. Gaussian and real-data are addressed separately
by training two different networks. The combination of CNN and $\F$-statistic results is done manually:
samples with high $\F$-statistic values are classified according to the $\F$-statistic; samples with
low $\F$-statistic values are classified according to the CNN.

Finally, \team{BearWaves}~\cite{third_team} presented a more involved approach. First, training data
was simulated on-the-fly using \texttt{PyFstat} to avoid overfitting. To simplify the task of combining
data from two detectors, data samples were time-averaged to obtain a common set of timsetamps. Data samples
were analyzed by an ensemble of neural networks applying different processing strategies.

Based on their relative leaderboard positions, image-based machine-learning strategies tend
to deliver less sensitive pipelines compared to track-based statistics unless a significant amount of resources
is invested in data pre- and post-processing, as shown by \team{BearWaves}~\cite{third_team}.

\subsection{Reference submission: \texttt{SOAP}\label{sec:soap}}

We now describe the application of \texttt{SOAP}~\cite{soap}
to the competition's dataset. \texttt{SOAP} is a fast, model-agnostic blind CW search capable
of detecting a broad class of CW sources, even if they are affected by complicated
physics such as spin-wandering, glitches, or timing noise~\cite{Carlin:2025sxm}.
This robustness, however, limits the sensitivity of \texttt{SOAP} to well-modelled isolated NSs
due the breadth of its parameter space~\cite{Wette:2023dom}.

Contrary to most blind CW searches, which identify interesting parameter‑space 
templates or regions where a signal may be present~\cite{Tenorio:2021njf},
\texttt{SOAP}’s main search stage assigns detection statistics to frequency \emph{bands} 
by returning the most significant track constructed via the Viterbi algorithm~\cite{viterbi}.
This matches the submission format of the Kaggle competition and makes \texttt{SOAP} a natural baseline
for comparing against Kaggle solutions.

The \texttt{SOAP} analysis begins 1800s SFT power spectra, which are summed over one-day 
periods (48 SFTs). This daily summation process effectively average out antenna 
pattern modulation and increasing the SNR in each frequency bin.
The search uses the line-aware statistic~\cite{soap}, 
a multi-detector approach that computes the Bayesian odds ratio between 
the signal hypothesis and the noise hypothesis including spectral 
lines~\cite{Covas:2019jqa}. This statistic specifically penalises instrumental 
line-like features in spectrograms, helping to distinguish genuine astrophysical 
signals from detector artifacts.
We use the same \texttt{SOAP} configuration as that used in~\cite{KAGRA:2022dwb}.
For each frequency band in the test dataset described in Sec.~\ref{sec:dataset},
a value of the line-aware statistic is produced for each band.

Note that \texttt{SOAP}'s signal model only requires continuity in the
frequency-evolution track across consecutive days.
No restrictions are set on the coherence of the signal, 
or the specific functional form of the frequency and amplitude modulation 
due to the detector's motion.
This freedom gives rise to the large parameter space covered by this search.
As a result, if a Kaggle solution surpasses the performance of \texttt{SOAP},
it suggests an enhanced capability to identify structures in the data 
that are characteristic of CW signals, likely because the method incorporates
more informative signal priors in its design.

\subsection{Comparison}

\begin{figure}
    \centering
    \includegraphics[width=\textwidth]{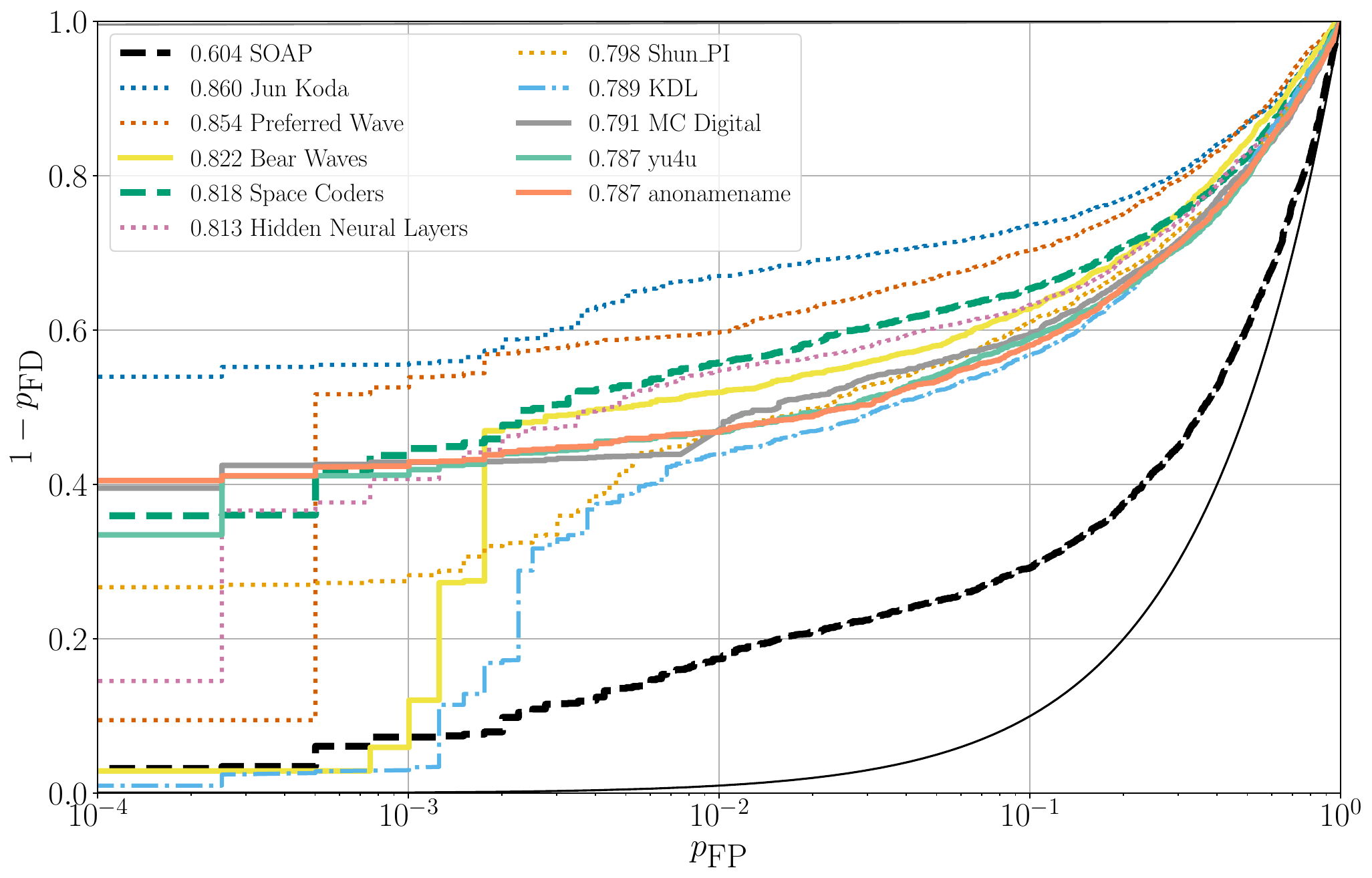}
    \caption{
        ROC curves corresponding to the top 10 solutions of the competition.
        The legend lists the AUC scores corresponding to each solution.
        The thin black line corresponds to a non-discerning classifier with an AUC score of 0.5.
    }
    \label{fig:final_roc}
\end{figure}

The performance of the competition's top-10 solutions is summarized using a ROC curve in
Fig.~\ref{fig:final_roc}, which shows the false-dismissal probability $\pfd$ of a pipeline
(or, equivalently, the detection probability $1 - \pfd$) as a function of false-positive
probability $\pfp$. The false-positive probability $\pfp$ is defined as the fraction of noise samples
that get classified as a signal; the false-dismissal probability $\pfd$ corresponds to the fraction
of signal samples that get classified as noise.

All solutions in the top 10 leaderboard produce an AUC above the value reported by
the \texttt{SOAP} reference submission in Sec.~\ref{sec:soap}. This indicates that all pipelines 
introduce cogent information into their assumptions when compared to an agnostic search,
as previously discussed.

The AUC score, however, aggregates results computed from the complete dataset,
which contains signals with different SNRS, across multiple false-positive probabilities.
As a result, there is no single solution that dominates across all false-positive probabilities. 
To more directly compare to standard CW search methods, we present results as a function of
SNR in Sec.~\ref{sec:cw_interpretation}.

\section{Comparison to CW searches\label{sec:cw_interpretation}}

Blind CW searches operate by identifying interesting regions in a broad
parameter-space to then follow them up with a more sensitive method or using a 
different dataset. The result of a search is a candidate
(or collection of candidates) with a relatively-low associated
significance that will be increased through follow-up 
steps if they are caused by CW signals. 
The solutions in this competition, due to its design, 
operate on data samples directly and return an associated 
false-positive probability $\pfp$;
that is, the significance they return is not associated to a specific template.
Hence, these two approaches are not directly comparable.

On the other hand, the false-dismissal probability $\pfd$,
which is defined as the fraction of signals not detected by a method,
is equivalent in both Kaggle solutions and CW searches, 
as in both cases it refers to the fraction of bands containing a signal 
that ended up labelled as noise. 

This equivalence allows us to interpret Kaggle solutions as one-shot CW searches. 
Specifically, we consider the situation in which a Kaggle solution is configured
to run a CW search at a relatively small $\pfd$ so that the Kaggle solution discards 
a fraction of about $(1 - \pfp)$ of the initial dataset, reducing the cost of the CW search, 
at a negligible sensitivity impact. The result of this first step would then be
followed-up using other methods, depending on the intent of the search.

The cost of a standard CW search can be modelled simply as
\begin{equation}
    \C_{\mathrm{CW}} = c_{\mathrm{t}} N_{\mathrm{t}}
\end{equation}
where $c_t$ is the average computing cost per template and $N_{\mathrm{t}}$ is the number of
templates evaluated by the search.  $N_{\mathrm{t}}$ itself can be further
expressed as 
\begin{equation}
    N_{\mathrm{t}} 
    = n_{\mathrm{t}} N_{\mathrm{bands}}
\end{equation}
where $n_{\mathrm{t}}$ is the average number of templates per band.
The cost of a search including a Kaggle-based pre-processing step, on the other hand,
can be expressed as 
\begin{equation}
    \C_{\mathrm{K} + \mathrm{CW}} 
    = c_{\mathrm{K}} N_{\mathrm{bands}} + \pfp \C_{\mathrm{CW}}
\end{equation}
where $c_\mathrm{K}$ is the cost of running the model in a band and $\pfp$ represents the fraction of bands
flagged as containing a signal by the method. 
The relative cost ends up as
\begin{equation}
    \frac{\C_{\mathrm{K} + \mathrm{CW}}}{\C_{\mathrm{CW}}} 
    = \pfp + q_{\mathrm{K}} .
\end{equation}
where $q_{\mathrm{K}} = c_{\mathrm{K}} / (c_{\mathrm{t}} n_{\mathrm{t}})$.
As a result, Kaggle solutions will provide an advantage if they are  computationally cheaper
($q_{\mathrm{K}} < 1$) and can reach a sufficiently low $\pfp$ at a low $\pfd$ level.

Strictly speaking, $c_t$ and $n_t$ depend on the specific implementation of a CW 
search~\cite{Wette:2023dom}. We here instead take an educated guess based on the latest result in the literature. 
The latest \eah{} searches conducted in LIGO O2 and O3 data~\cite{eah_O2, Steltner:2023cfk} use
$n_{\mathrm{t}} \sim 10^{12} - 10^{15}$, depending on the frequency band. Similarly, 
the computing cost of the $\F$-statistic is about 
$c_\mathrm{{t}} \sim(10-100)\, \mathrm{ns}$~\cite{Dunn:2022gai,eah_O2, Steltner:2023cfk, Wette:2021tbv}.
Thus,  $c_{\mathrm{t}} n_{\mathrm{t}} \sim (10^4 -10^7) \, \mathrm{s}$.

The computing cost per-band is reported by some of the participants:
\team{JunKoda} reports a cost of $5$ days on a GPU to evaluate the full dataset 
($c_{\mathrm{K}} \approx \SI{50}{\second}$); \team{PreferredWave} reports $2$ 
days on a GPU ($c_{\mathrm{K}} \approx \SI{30}{\second})$; 
\team{Hidden Neural Layers} reports $10$ hours
using 8 CPU cores ($c_{\mathrm{K}} \approx \SI{5}{\second}$); 
\team{Shun\_PI} reports about 1 day using 1 CPU ($c_{\mathrm{K}} \approx \SI{1}{\second}$).
No computing cost estimation is available for the rest of the teams. For algorithms based on numerical
optimisation we conservatively assume $c_{\mathrm{K}} \approx (1 -10)\, \mathrm{s}$, while 
those based on neural networks can reach $c_{\mathrm{K}} \approx (0.1 -1)\, \mathrm{s}$ if evaluations
are batched on a GPU. Overall, we find $q_{\mathrm{K}} \lesssim 10^{-3}$.

\begin{figure}
    \includegraphics[width=\textwidth]{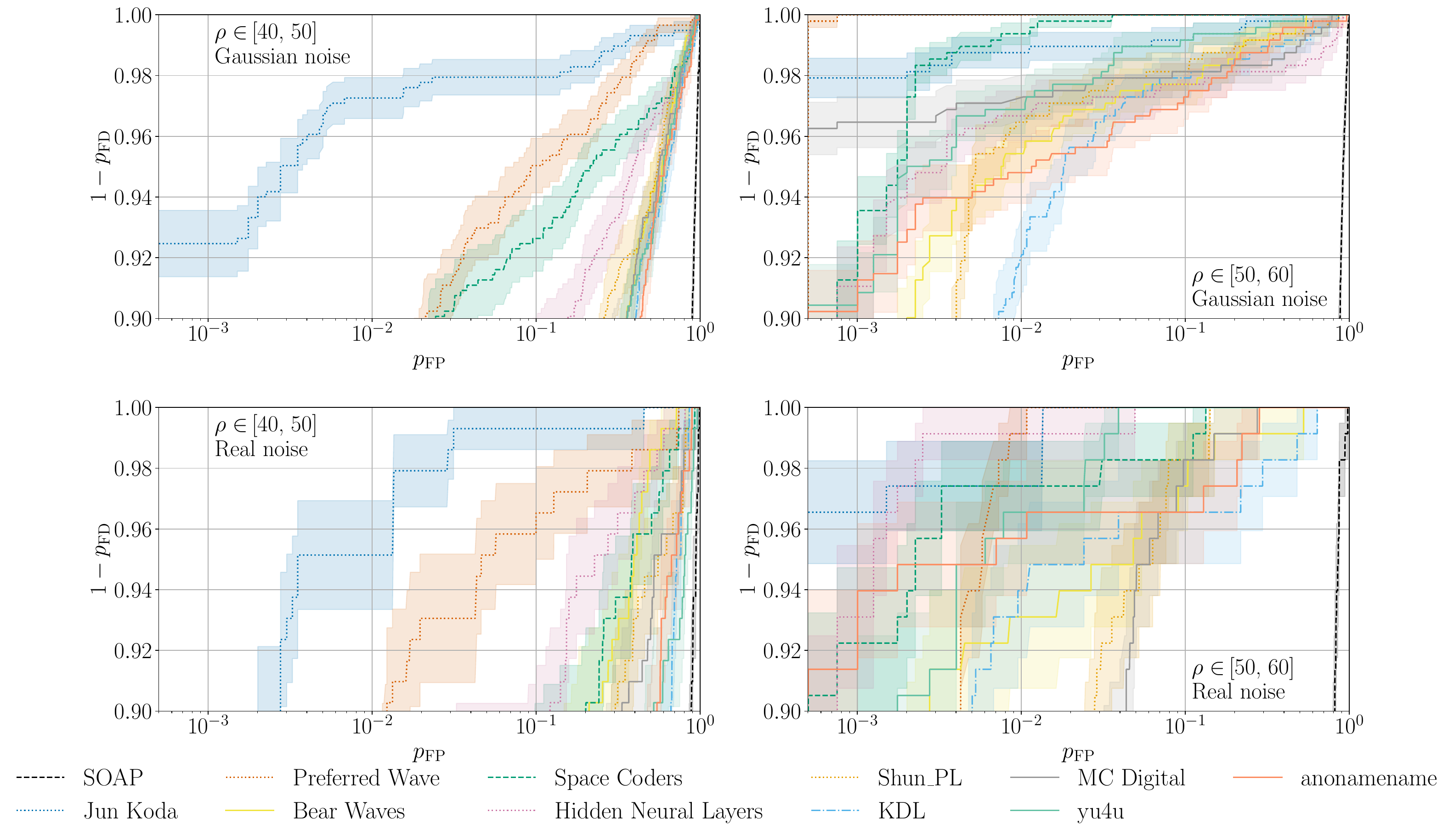}
    \caption{
        False dismissal probability as a function of false-positive probability 
        for a restricted range of SNR. 
        The upper row shows the results using Gaussian noise samples,
        while the lower row uses only real-data samples.
        The shaded regions show binominal errorbars
        $\delta \pfd = \sqrt{\pfd  (1 - \pfd)/N_{\mathrm{s}}}$,
        where $N_{\mathrm{s}}$ is the number of signals in said SNR range.
    }    
    \label{fig:pfp_vs_snr}
\end{figure}

We finally evaluate the performance of each Kaggle solution in restricted SNR ranges matching the two 
reference searches previously discussed, namely $\rho \in [40, 50]$ corresponding to $\weave{}$ and
$\rho \in [50, 60]$ corresponding to $\eah{}$. The results are shown in Fig.~\ref{fig:pfp_vs_snr} for Gaussian
and real-data samples. The number of signal samples within said ranges limits the resolution of 
false dismissal probabilities to about $(1-2)\%$, and consequently we evaluate $\pfp$ at $1 - \pfd = 0.98$.
Similarly, the precision on $\pfp$ is about $10^{-3}$; given the computed value of $q_{\mathrm{K}}$, this implies
the computational gains are dominated by $\pfp^{-1}$.

For the \weave{} range ($\rho \in [40, 50]$), 
in the case of Gaussian noise, \team{JunKoda}
reaches a false dismissal probability $\pfd = 0.02$ at $\pfp \sim (10^{-2} - 10^{-1})$, 
while \team{Preferred Wave} performs at $\pfp \sim (0.2 - 0.4)$. All other solutions
achieved the allowed false dismissal probability at $\pfp \gtrsim  0.5$. The potential gain
in this case is from a factor of about 5 (\team{Preferred Wave}) to one to two orders of magnitude 
(\team{JunKoda}). The situation in the case of real noise remains comparable within the uncertainty.

For the \eah{} range ($\rho \in [50, 60]$), in Gaussian noise, \team{JunKoda}, \team{Preferred Wave}
and \team{Space Coders} reach $\pfp \sim 10^{-3}$ at the specified false dismissal probability level. 
The next relevant team is \team{yu4u} (9th), at $\pfp \sim 3 \times 10^{-2}$.
The remaining solutions about $\pfp \sim 10^{-1}$. 

Despite the increased uncertainty, the situation in real-data samples changes slightly: 
\team{JunKoda} stays at $\pfp \sim 10^{-3}$;  \team{Preferred Wave} worsens up to $\pfp \sim 7 \times 10^{-3}$,
and \team{Space Coders} loses an order of magnitude up to $\pfp \sim 3 \times 10^{-2}$. \team{yu4u} maintains
a comparable performance to the Gaussian case.
The relative loss of \team{PreferredWave} with respect to \team{JunKoda} can be ascribed to the use of
average power in a template bank: In non-Gaussian data, the average template-bank power tends to rise
due to the presence of extreme values in the background, even if most of the problematic bins have been removed. 
Similarly, the loss of \team{Space Coders} can be related to the fact that their method builds-up a 
frequency-evolution track in a data-driven manner, an thus can get biased due to local maxima in the spectrogram.
Finally, we highlight the performance of \team{Hidden Neural Layers} (5th), which reaches $\pfp \sim 10^{-3}$ 
surpassing all other teams except \team{JunKoda}. This is likely a consequence of their approach to isolate
and remove potential glitches~\cite{fifth_team}.

\section{Discussion\label{sec:discussion}}

CW signals are one of the potential next discoveries of gravitational-wave astronomy.
Currently their detection is complicated by our ignorance on the specific emission
mechanism at play, which gives rise to a broad parameter-space that requires
significant computing budgets to be effectively covered. Fundamentally, the
sensitivity of CW searches for unknown sources is limited by the available 
computing~\cite{Prix:2012yu, Wette:2018bhc, Wette:2023dom}.

In this work, we explored the use of data-analysis competitions to discover new search strategies.
Specifically, we framed CW searches as a classification problem and deployed a Kaggle competition to
find the most effective strategy to identify data samples harbouring simulated CW signals.
Kaggle~\cite{kaggle} is a online data-analysis competition platform powered by Google hosting a
broad community of data-analysis practitioners with a wide variety of backgrounds. By tapping into a new, broad community of experts, 
we intended to asses the performance of
data-analysis methods from other field 
when applied to the problem of detecting CW signals. 
This approach is similar to mock data challenges previously developed for other
gravitational-wave use cases~\cite{Messenger:2015tga,Walsh:2016hyc,Baghi:2022ucj,Schafer:2022dxv}.

The challenge lasted for 3 months and attracted more than 1,000 participants. 
The specific task was to classify 8,000 data samples (containing Gaussian noise, 
real detector data, and possibly simulated CW signals).
Each submission was ranked according to the Area under the ROC curve metric.
Upon completion, we contacted the top 10 solutions to understand their 
approach and implementation.

Overall, the competition was dominated by pipelines based on
semicoherent matched filtering, in a similar fashion
to~\cite{Krishnan:2004sv,2014PhRvD..90d2002A}. 
The proposed solutions, however, introduced several improvements
with respect to well established methods. 

First, the evaluation of  detection statistics was vectorised over 
multiple waveforms using GPUs without using any signal approximation.
This contrasts with most GW GPU-accelerated methods so far,
which either parallelize the evaluation of a single 
template~\cite{Dunn:2022gai, Garcia-Quiros:2025usi}
or compute an approximated version of the detection 
statistic~\cite{Covas:2019jqa, Rosa:2021ptb}.
The potential improvement in computational cost and complexity of search pipelines 
inspired the development of \texttt{fasttracks}~\cite{Tenorio:2024jgc}, a unified, 
GPU-accelerated engine to compute CW detection statistics using SFTs for general 
CW waveform models. 

Second, the winning solution~\cite{first_team} made use of alternative
filtering strategies, similar to those previously outlined in Ref.~\cite{Allen:2002bp}.
While these specific statistics have not been applied to short-coherence 
CW searches, their efficiency allowed us to extend the use-case
of efficient SFT-based detection statistics to the search for
binary neutron stars and massive black-hole binaries~\cite{Tenorio:2025gci}
in next-generation ground-based detectors (Einstein Telescope~\cite{Abac:2025saz}, 
Cosmic Explorer~\cite{Reitze:2019iox}). Similar strategies have been proposed
to analyze GW signals in the LISA~\cite{LISA:2024hlh} frequency band, such as galactic
binaries~\cite{Prix:2007zh, Blaut:2009si}, which are modelled in a similar way to LIGO 
CW sources, and more recently a broader class of systems such as stellar-mass binary
black holes~\cite{Bandopadhyay:2025fyx} and extreme mass-ratio inspirals~\cite{Speri:2025ucn}.

Solutions accounting for the presence of non-Gaussian data did so in a relatively 
simple manner, either by masking problematic regions in the 
spectrogram~\cite{first_team, second_team, fifth_team, sixth_team}
or by generating non-Gaussian training 
data~\cite{third_team, eighth_team, nineth_team, tenth_team}. 
This is comparable to the practical approach taken in CW searches, 
where non-Gaussianities can be identified as sudden spikes in the accumulated 
power~\cite{Covas:2019jqa, Tenorio:2024jgc}.

Our analysis suggests that, for datasets compared to those used in this
competition, the use of Kaggle solutions could reduce the number of 
frequency bands to  be evaluated by one to two orders of magnitude. 

The relevance of approaches based on alternative techniques, such as machine learning, 
becomes apparent when analyzing  specific subsets of SNR.
For example, for moderately strong signals $\rho \in [50, 60]$, 
we find that U-net architectures~\cite{nineth_team} are able to compete with 
track-based statistics and show a negligible performance 
degradation when applied to non-Gaussian data. 

Despite the limited duration of this challenge, the obtained solutions have the
potential to affect future CW searches not only by improving the efficiency
and deployment of track-based searches~\cite{Tenorio:2024jgc,Tenorio:2025gci},
but also by spurring approaches prescinding from template banks all together.
This latter point is particularly interesting, as the computing cost of a CW search
is dominated by the evaluation of statistics on a template bank to numerically
marginalise with respect to the signal parameters to compute a 
Bayes factor~\cite{Prix:2009tq,Searle:2008jv,Prix:2011qv}; 
this Kaggle competition, on the other hand, allows for pipelines operating 
directly on data samples, so that the method itself constitutes an approximation
of the marginalised Bayes factor. The solutions here collected, together with those
proposed from within the community~\cite{Dreissigacker:2019edy,
Miller:2019jtp,Bayley:2020zfa,Yamamoto:2020pus,Bayley:2022hkz,
Modafferi:2023nzt,Joshi:2025xdz}, bear witness of the potential of alternative 
approaches to revolutionize the search for blind CW sources. 

Although the dataset is shorter than typical production searches, 
the chosen duration (4 months) remains a valid regime for candidate generation. 
Given the expected  cadence of forthcoming LVK observing runs, 
these results support a strategy based on analysing each run shallowly
to produce candidate lists, with significance subsequently accumulated across 
runs, akin to the “fresh-data mode” discussed in Ref.~\cite{Cutler:2005pn}. 
Such implementation will be presented elsewhere.

This Kaggle competition is the first public, large‑scale exposure of the 
CW detection problem to a wide community of data‑analysis practitioners, 
and a logical step beyond established closed‑door mock‑data 
challenges~\cite{Messenger:2015tga, Walsh:2016hyc, Baghi:2022ucj, Schafer:2022dxv} 
to foster cross‑pollination across fields. 
Its impact is already evident in new GPU‑accelerated
methods~\cite{Tenorio:2024jgc} and in applying CW‑inspired detection statistics
to other domains, such as compact binary coalescenses in next-generation 
GW detectors~\cite{Tenorio:2025gci,Speri:2025ucn,Bandopadhyay:2025fyx}. 
The materials developed for this competition, including tutorials for generating 
and pre‑processing CW data with \texttt{PyFstat}~\cite{pyfstat}, 
and the dataset itself~\cite{dataset}, are publicly available to support
further development of alternative methods aimed at the first CW detection.

\section*{Acknowledgements}

We are grateful to the all Kaggle competitors for their invaluable input and engagement 
throughout the competition.
We thank
Rafel Jaume, David Keitel, Narenraju Nagarajan, Alicia M. Sintes, and Daniel Williams
for support and discussions,
and Christopher Berry, Lorenzo Mirasola, and Karl Wette for comments on the manuscript.
We thank Kaggle/Google for contributing to the competition prize,
and acknowledge support from COST Action CA17137 (G2Net).
RT thanks Graham Woan and the Institute for Gravitational Research at 
the University of Glasgow for their hospitality during the development of this work.
RT is supported by 
ERC Starting Grant No. 945155–GWmining;
Cariplo Foundation Grant No. 2021-0555;
MUR PRIN Grant No. 2022-Z9X4XS;
MUR Grant “Progetto Dipartimenti di Eccellenza 2023-2027” (BiCoQ);
Italian-French University (UIF/UFI) Grant No.~2025-C3-386;
the ICSC National Research Centre funded by NextGenerationEU;
the Universitat de les Illes Balears (UIB);
the Spanish Agencia Estatal de Investigación grants PID2022-138626NB-I00, RED2024-153978-E, RED2024-153735-E,
funded by MICIU/AEI/10.13039/501100011033 and the ERDF/EU; 
and the Comunitat Autònoma de les Illes Balears through the 
Conselleria d'Educació i Universitats with funds from the 
European Union - NextGenerationEU/PRTR-C17.I1 (SINCO2022/6719) 
and from the European Union - European Regional Development Fund (ERDF) (SINCO2022/18146).
MJW is supported by the Science and Technology Facilities Council [2285031, ST/X002225/1, ST/Y004876/1], 
 and the University of Portsmouth. JB and CM are supported by STFC grant ST/V005634/1.
Computational work was performed at MareNostrum5 with technical support provided by 
the Barcelona Supercomputing Center (RES-FI-2024-3-0013, RES-FI-2025-1-0022),
at CINECA with allocations through INFN and Bicocca.
This research has made use of data or software obtained from the Gravitational Wave 
Open Science Center (gw-openscience.org), a service of LIGO Laboratory, 
the LIGO Scientific Collaboration, the Virgo Collaboration, and KAGRA.
This paper has been assigned document number LIGO-P2200295.

\appendix

\section{First Kaggle Competition\label{sec:first_kaggle}}

The G2Net Kaggle challenge (G2Net Gravitational Wave Detection), launched on 30 June 2021, marked the first public data competition in gravitational-wave astrophysics aimed specifically at the machine learning community. Its goal was to explore the capabilities of machine learning classifiers for detecting transient GW signals from stellar-mass binary black hole (BBH) mergers. Competitors were tasked with predicting the presence or absence of a BBH signal within short time-series recordings of simulated detector data. Similarly to the CW challenge discussed in this paper, the performance was also evaluated using the AUC, computed from participants’ probability estimates of signal presence in the test set.

The dataset comprised simulated measurements from the three-detector network of advanced interferometers: LIGO-Hanford, LIGO-Livingston, and Virgo. Each training example consisted of three time series, corresponding to the three detectors, covering two seconds sampled at 2048 Hz. Some samples contained only Gaussian noise, while others were augmented with simulated GW signals. The simulated BBH signals were generated using \texttt{IMRPhenomPv2}~\cite{Hannam:2013oca} and varied across 15 astrophysically relevant parameters, including component masses, spins, binary orientation, sky location, distance, polarization, merger phase, and arrival time. Although the distributions of these parameters reflected realistic astrophysical populations, the overall signal amplitudes were adjusted to produce a mix of challenging and easier detection cases. Most signals were designed to lie near the typical sensitivity threshold of matched-filter searches, around an SNR of 8.

The challenge attracted broad participation, with 1,501 individuals forming 1,219 teams. Competitors applied a variety of machine learning approaches, from convolutional and recurrent neural networks to hybrid architectures, often incorporating data augmentation and ensemble methods to improve generalization. The competition validated claims in the literature~\cite{Gabbard:2017lja,2018PhRvD..97d4039G,2019PhRvD.100f3015G} that deep learning methods could achieve competitive detection performance on realistically simulated multi-detector GW datasets. Moreover, the challenge provided a benchmark for future efforts in applying modern machine learning techniques to gravitational-wave detection, highlighting both the potential and limitations of data-driven approaches in this domain.

\begin{figure}
    \centering
    \includegraphics[width=\textwidth]{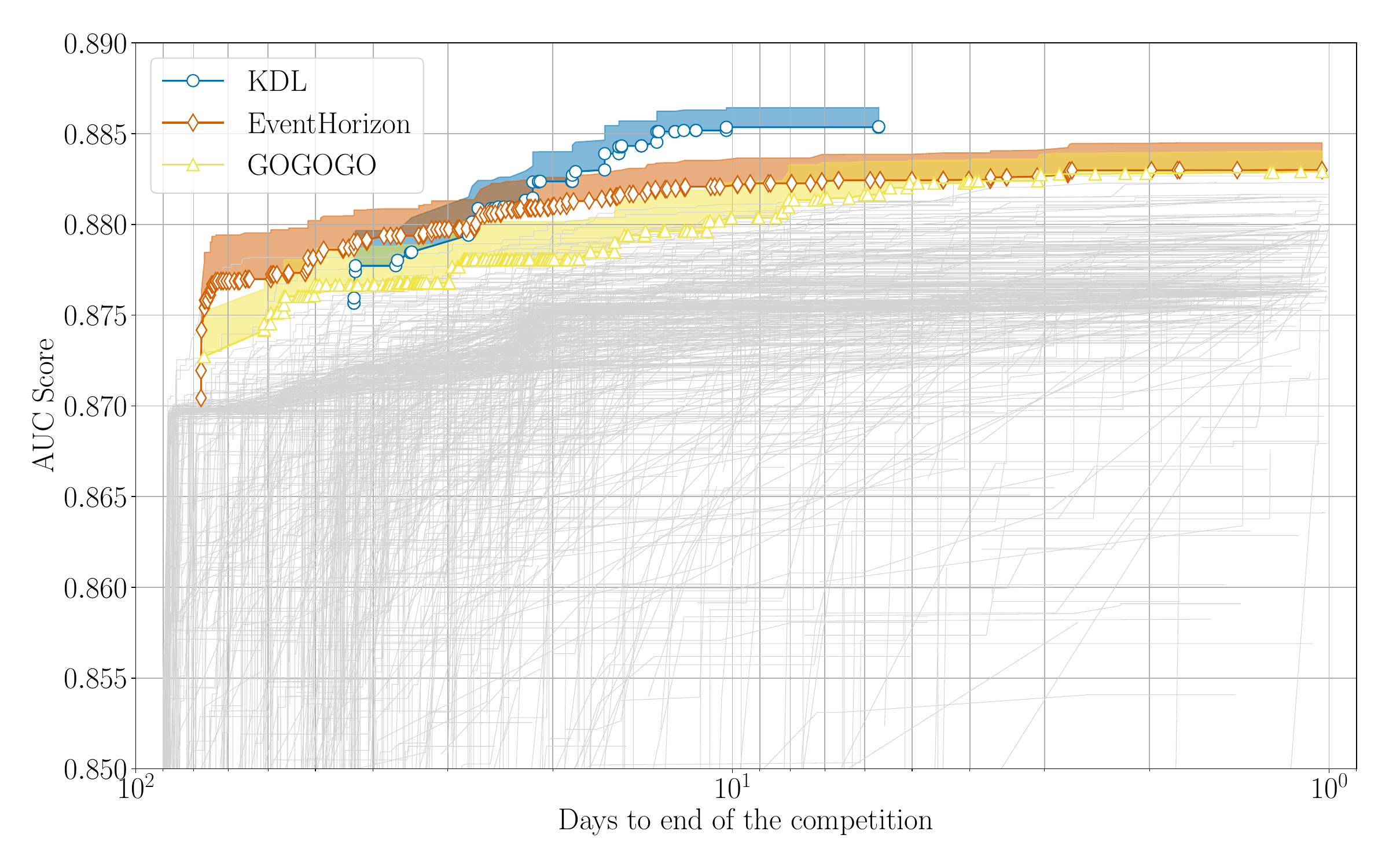}
    \caption{Evolution of the top-3 teams throughout the competition in terms the AUC score. 
        The markers correspond to the private leaderboard score, while the shaded regions
        extend to the public leaderboard score. The evolution of the private leaderboard score for other teams is shown in grey.}
    \label{fig:BBH_evolution}
\end{figure}

\begin{figure}
    \centering
    \includegraphics[width=0.48\textwidth]{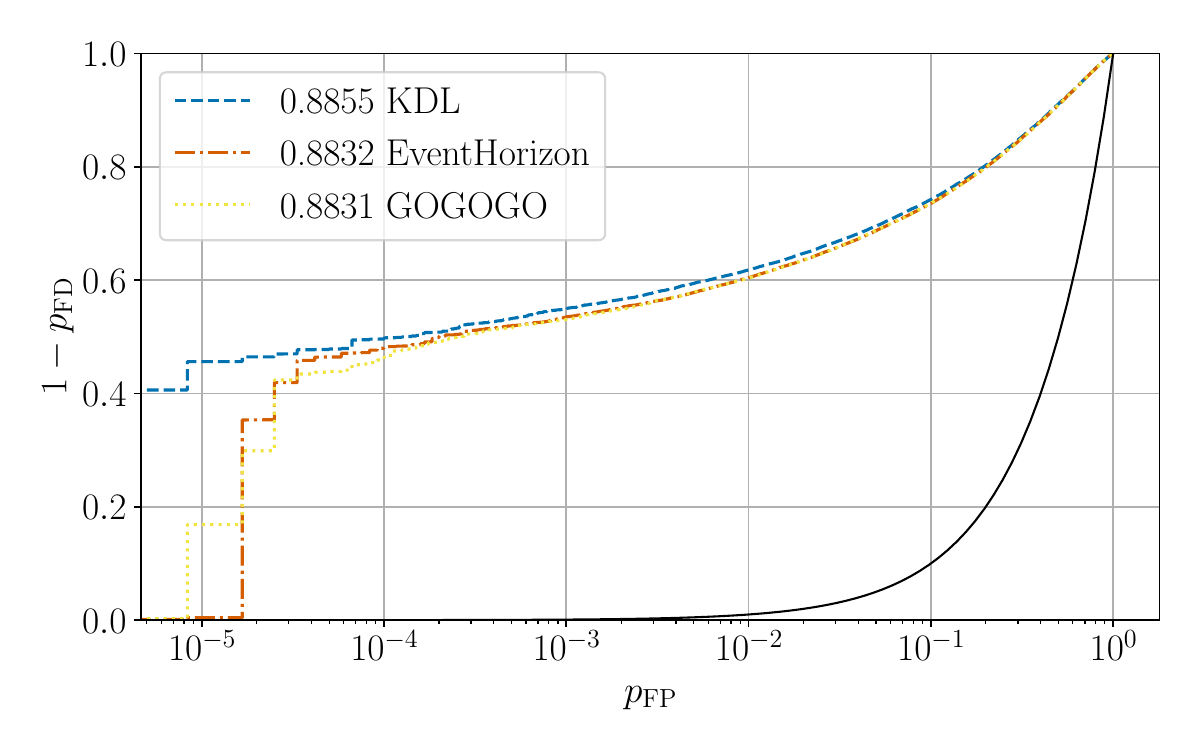}
    \includegraphics[width=0.48\textwidth]{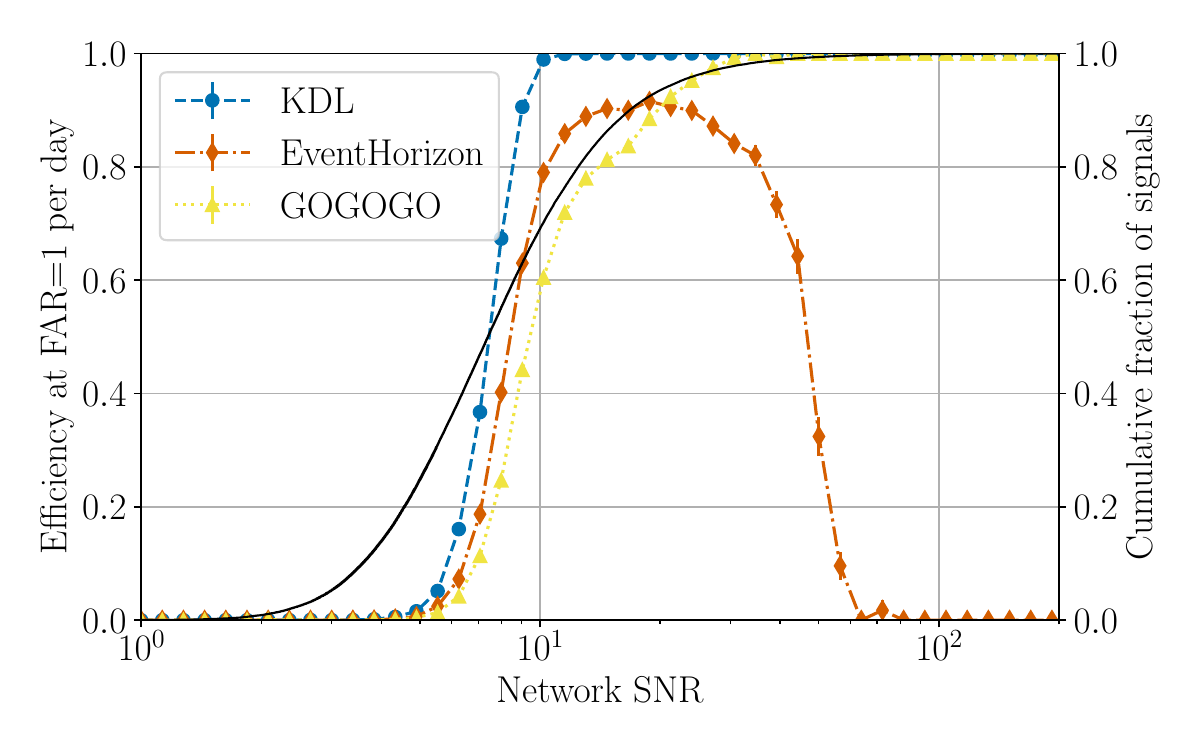}
    \caption{
        \textbf{Left:} ROC curves corresponding to the top-3 solutions of the first Kaggle competition. The legend lists the AUC scores corresponding to each solution. The black line corresponds to a non-discerning classifier with an AUC score of 0.5.
        \textbf{Right:} Efficiency as a function of network matched SNR (computed using the injection parameters) at a FAR of 1 per day for the top-3 solutions of the first Kaggle competition. We use 50 bins with logarithmic spacing between SNR 1 and 100 to produce the curve. The black line shows the cumulative fraction of signals in the original dataset as a function of SNR.
    }
    \label{fig:BBH_ROC}
\end{figure}

In Fig.~\ref{fig:BBH_evolution}, we show the evolution of the AUC scores achieved by all participating teams over the course of the competition. Most final solutions clustered tightly around an AUC of $\sim 0.88$. The highlighted curves for the top three solutions indicate that their initial approaches could readily achieve AUCs of $\sim 0.875$. The ROC curves in Fig.~\ref{fig:BBH_ROC} provide additional insight into the relative performance of the top three solutions. Of particular note is that the winners, \team{KDL}, were able to gain an advantage over 2nd place \team{EventHorizon} and 3rd place \team{GOGOGO}, achieving marginally higher overall probability of detection but, importantly, substantially better performance at the lowest false-positive rates. Figure~\ref{fig:BBH_ROC} also provides the detection efficiencies of each of the top three teams as a function of network SNR at a fixed false positive probability equivalent to 1 per day. It is interesting to see clear distinction in sensitivity between the teams at efficiencies $\sim 0.5$ but the KDL team achieve efficiencies of 1 at significantly lower SNR. We also note that the 2nd place efficiency curve indicates a potential dataset bias or  extrapolation failure where under-representation in training can lead to extremely loud signals being outside the learned feature distribution and hence classified incorrectly as noise.

The top three teams share several key strategies that contributed to their performance. They emphasised careful preprocessing of the waveform, including whitening, bandpass filtering, and mitigation of edge effects, while selectively using augmentation techniques such as Gaussian noise, waveform flipping, and time shifts to improve generalisation. Each team used 1D convolutional architectures as primary frontends, often in combination with 2D CNN backends or hybrid models, and recognised the importance of channel-specific processing—sharing weights for similar detectors while handling others separately. Synthetic data generation and pre-training played a critical role in avoiding overfitting and providing access to auxiliary targets, and all groups used ensembles of multiple models to further boost robustness. Finally, iterative approaches like pseudo-labeling, test-time augmentation, and progressive refinement were widely employed to exploit unlabelled data and maximize leaderboard performance.

The winning team (KDL) overcame the main competition challenge — overfitting— by generating large amounts of realistic synthetic data for pre-training, which also allowed the use of hidden parameters as auxiliary targets. Their model design included learnable frontends for each channel, transformation of 1D data into richer time-frequency or feature spaces, and lean encoders to limit overfitting.
Key innovations included a custom 1D CNN with large kernels, recognition that 1D models outperform 2D CNNs, and extensive use of synthetic data for pre-training. They also observed fundamental limitations: that about 30\% of true signals are inherently undetectable, some noise samples are always misclassified as signals, and training on difficult samples can invert model predictions.


\printbibliography

\end{document}